\documentclass{article}
\usepackage{spconf,amsmath,graphicx}

\usepackage{enumitem}
\usepackage{makecell}
\setlist{nosep, leftmargin=14pt}

\usepackage{mwe} 
\usepackage{color}
\usepackage{hyperref}
\newcommand{\etal}{\textit{et al. }}

\title{Deep learning Framework for Mobile Microscopy}
\name{\makecell{Anatasiia Kornilova, Mikhail Salnikov, Olga Novitskaya, Maria Begicheva, \\ Egor Sevriugov, Kirill Shcherbakov, Valeriya Pronina, and Dmitry V. Dylov}}
\address{Skolkovo Institute of Science and Technology, 30/1 Bolshoi blvd., Moscow, 121205 Russia}
\begin{document}
%
\maketitle
\begin{abstract}
{Mobile microscopy is a promising technology to assist and to accelerate disease diagnostics, with its widespread adoption being hindered by the mediocre quality of acquired images.  Although some paired image translation and super-resolution approaches for mobile microscopy have emerged, a set of essential challenges, necessary for automating it in a high-throughput setting, still await to be addressed. The issues like in-focus/out-of-focus classification, fast scanning deblurring, focus-stacking, \textit{etc.} -- all have specific peculiarities when the data are recorded using a mobile device. In this work, we aspire to create a comprehensive pipeline by connecting a set of methods purposely tuned to mobile microscopy: (1) a CNN model for stable in-focus / out-of-focus classification, (2) modified DeblurGAN architecture for image deblurring, (3) FuseGAN model for combining in-focus parts from multiple images to boost the detail. We discuss the limitations of the existing solutions developed for professional clinical microscopes, propose corresponding improvements, and compare to the other state-of-the-art mobile analytics solutions.}
\end{abstract}



\section{Introduction}
\label{sec:intro}

Mobile microscopy is a modality that uses a smartphone camera and computational resources as a powerful tool for fast disease detection and improvement of at-home diagnostics and telemedicine level. Distinctive features of mobile microscopy are artifacts in the optical system, such as dust, water drops, condensation, and lower image quality compared to professional clinical microscopes due to the lower camera matrix quality and cheaper optics. Because of these traits standard microscopy image processing techniques cannot be directly applied to images acquired with a mobile microscope. 

Biomedical microscopy image processing involves the following levels of algorithms: (1) low-level algorithms, such as filtering of irrelevant images and common enhancement of image quality, (2) high-level algorithms that use semantic information from an image to perform segmentation, cell counting, and disease detection. Majority of relevant works on image processing for mobile microscopy is devoted to the second type of algorithms, namely quantification of blood-borne filarial parasites~\cite{Ambrosio2015}, malaria~\cite{Pirnstill2015} and tuberculosis diagnosis~\cite{Tapley2013}. Only a few works are focused on image quality enhancement via supervised deconvolution with clinical microscope~\cite{Rivenson2018_2}, whereas other important subtopics are crucial in processing massive amounts of data.

To fill this gap, in our work we present solutions to the following problems in the application to mobile microscopy:
\begin{itemize}
    \item \textbf{In-focus/out-of-focus images classification.} Despite the presence of auto-focusing systems in modern microscopes acquired images still must be filtered to remove blurred images with artefacts and to detect focal planes within the same sample in which different focus areas are present.
    \item \textbf{Fast scanning image deblurring.} Z-axis scanning usually takes much time to produce high quality images. This can be improved by reusing deep learning deblurring techniques for fast scanning, restoring an original image from the one acquired from a fast motion. 
    \item \textbf{Focus-stacking (Multi-Focus images fusion).} Since optical microscopes usually have a shallow depth of field, a volume sample cannot be examined from a single focal length. To understand the structure of a sample one must examine several focal planes of a sample in which different focus areas are located. One popular technique to do this is focus-stacking or multi-focus image fusion.
\end{itemize}

\vspace{.2cm} 
Deep learning is a powerful family of machine learning methods that successfully address a wide range of biomedical applications, in particular microscopy image processing tasks such as classification, segmentation, super-resolution, deblurring and denoising \cite{Xing2018}, which are also relevant in mobile microscopy. Thus, in our work we propose to apply deep learning image processing techniques to solve considered mobile microscopy problems. To the best of our knowledge, this is the first work where the mobile microscopy image processing problems are tackled using deep learning techniques.

In this paper, we present the deep learning framework for mobile microscopy, a set of techniques that addresses the following fundamental problems of mobile microscopy imaging: in-focus/out-of-focus classification, fast scanning deblurring, and focus-stacking. Images from a smartphone can be sent to a server where they will be processed by a pretrained deep learning model. We emphasize why existing methods, both classical and deep learning, cannot be directly applied to mobile microscopy data, and suggest modifications to address these problems. As a result, we obtain a set of effective methods that help to significantly speed up data processing and improve image quality for such a promising field as mobile microscopy.



\section{Methods}
\label{sec:methods}

We use a z-stack video dataset of 150 specimens from the auto-scanning mobile microscope with the most generic configuration, both in terms of its price and the ease of use, provided by MEL Science \cite{MEL} for all the experiments. Validation on other microscope setups will be the scope of future work. All samples are taken from the standard slide sets manufactured by AmScope and recorded at four scan speed levels in the z-axis.

The experiments are conducted on Intel Xeon 2.20GHz CPU and Tesla T4 GPU using PyTorch, NumPy and OpenCV libraries.

\subsection{In-focus/out-of-focus images classification}

The stage of data pre-processing from mobile microscopes to filter focused images is one of the most critical stages as it can significantly save the healthcare professional's time in processing samples. The classical solution to perform such a pre-processing step is to measure image sharpness using focus measures operators based on Laplacian or Sobel operators. Yet, this approach is sensitive to distortions on sample borders and tends to produce many false positives.

\begin{figure}[htb]
\centering
\centerline{\includegraphics[width=0.8\columnwidth]{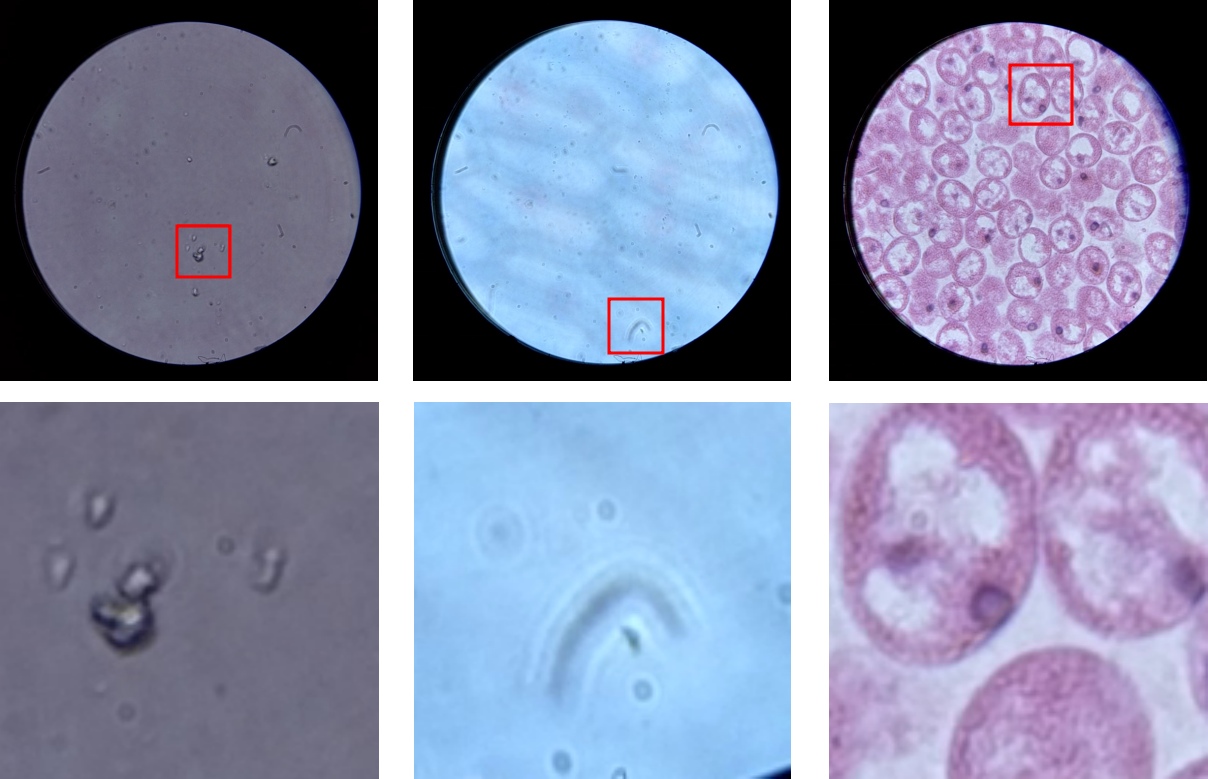}}
\caption{Examples of images with artifacts to be filtered out: condensate (left) and dust in optical system (center); example of a correct image with focused specimen (right).}
\label{fig:in-focus}
\end{figure}

To overcome this hurdle, many authors, including~\cite{Senaras2018, Kohlberger2019}, turned to DL architectures and formulated the problems of distinguishing out-of-focus images as a classification task. Nevertheless, these networks are trained and adapted to a dark-field fluorescence microscopy data and cannot be easily applied to the case of mobile microscopy. Another important problem related to mobile microscopes is the presence of different artifacts caused by cheap optics and less thorough specimen recording, e.g. dusty images or images with condensate (Fig.~\ref{fig:in-focus}).

In our work we propose to implement the approach based on the work~\cite{Yang2019}, in which authors consider the multi-label classification, each label refers to a certain level of defocus. To obtain different levels of defocus for synthetic defocusing of images the authors use the mathematical model of point spread function, based on Bessel function. Since our dataset contains the original z-stack of the microscope we use original images from the stack with a constant step over the z-axis instead of synthetic data. The final network architecture contains two concatenated convolutions layers with max pooling and ReLU activation and two fully connected layers. Another important feature of the model proposed in our work is inclusion of a rank probability score for multi-label classification instead of cross-entropy that shows better results for our task.

In order not to lose frames with small focused areas from the image stream, we consider cropping images to a constant size instead of using the entire image. We empirically found that cropping to size $84 \times 84$ and use of 10 focus levels, i.e., classes, provide the best performance of our network. We train our network with 1500 epochs and optimize it with Adam optimizer with $1 \cdot 10^{-6}$ learning rate; the batch size is equal to 8, cross-entropy is used as a loss function.

\subsection{Fast scanning image deblurring}
In mobile microscopy algorithmic solutions for obtaining sharp images from blurred ones allow the use of cheaper equipment or fast scanning techniques. 



Recently convolutional neural networks (CNN) proved to be an effective tool for the task of image deblurring. In~\cite{Dong2015} Chao Dong \etal proposed SRCNN architecture for high resolution imaging, which directly learns a mapping function between low and high resolution images. This architecture, consisting of three convolution layers and two ReLU activation functions between, can be applied to image deblurring. Then generative adversarial networks (GANs) became a popular instrument for solving super resolution~\cite{Ledig2017} and image deblurring problems. In 2018 Orest Kupyn \etal proposed DeblurGAN~\cite{Kupyn2018}, a powerful architecture based on a conditional GAN and the content loss, allowing to achieve a great performance on motion blur dataset.

In our work we propose to employ the aforementioned architectures for the fast scanning image deblurring task. We start with exploring the classical U-net architecture~\cite{Ronneberger2015} and SRCNN~\cite{Dong2015}. We train both models for 450 epochs with learning rate $10^{-3}$ of Adam optimizer. 

Then we examine a more complex model developed specifically for the task of image beblurring -- DeblurGAN, whose standard implementation is based on the original article (sDeblurGAN)~\cite{Kupyn2018}. We propose upgraded DeblurGAN (uDeblurGAN) obtained by replacing the 2D transpose convolution generator layers resize-convolution layers to avoid producing images with color checkerboard artifact pattern by generator~\cite{Odena2016}. The uDeblurGAN generator was pre-trained on input images to avoid discriminator overfitting on the noise produced by generator.



To train all models we select 192 pairs of blurred and ground truth images from the microscope scanning. For faster computations we use cropping and resizing of images to the size of $256\times256$.

For the uDeblurGAN generator pre-training we use data augmentation techniques such as random horizontal flip, random vertical flip and random rotation on small images ($64\times64$). The entire uDeblurGAN training is conducted on images of size $256\times256$. We train all DeblurGAN-based models for 200 epochs with Adam optimizer for generator and discriminator with learning rate $10^{-4}$ and $\beta_1,\beta_2=(0.5, 0.999)$.

\subsection{Focus-stacking}
Recently DL-based focus-stacking methods for obtaining multi-focus images showed better performance over classical computer vision approaches (wavelet transforms, as most effective representative of that class) thanks to their ability to produce smoother images than the latter. Among DL-based methods the following  CNN-based models are widely known: CNN based~\cite{CNNbased}, IFCNN~\cite{IFCNN}, MLFCNN, MFNet. However, almost all of the mentioned models require presence of reliable ground truth images. Moreover, CNN based approaches tend to provide not very clear textures and low depth of field~\cite{Genga2020}. Recently the FuseGAN model, which has been shown to be effective in solving the above problems, was proposed in~\cite{FuseGAN}. The main advantage of GAN is the ability to train the generator while optimizing the focus-stacking error (L1Loss) as well as smoothness error (BCELoss). Later this model was improved and applied to biomedical imaging~\cite{Genga2020}.
However, in order to achieve the best image quality, it is necessary to select the most suitable loss function as well as hyperparameters. In the current section we explore this problem in the application of the FuseGAN for mobile microscopy.

The model structure consists of a generator, a discriminator and a model for generating masks. The generator and discriminator architectures are based on those proposed in~\cite{Genga2020}. To generate the masks we use algorithm based on the Modified Harris Corner Response Measurement~\cite{Sigdel2016}, which shows higher quality of the resulting masks as opposite to the neural network proposed in the original paper. 

A standard form of the loss function is used to train the discriminator. Due to the lack of ready-made ground truth images, images obtained using the IFCNN model~\cite{IFCNN} are used.

To ensure the best performance of the proposed model, we explore next combinations for generator loss function:
\begin{enumerate}
    \item \textit{BCELoss} + \textit{L1Loss} + \textit{Perceptual loss} (VGG16),
    \item \textit{BCELoss} + \textit{L1Loss} + \textit{Perceptual loss} (discriminator),
    \item \textit{BCELoss} + \textit{uL1Loss} + \textit{Perceptual loss} (discriminator).
\end{enumerate}
Here \text{L1Loss} refers to the absolute error between the model output and the ground truth, \textit{Perceptual loss} is calculated on hidden outputs of VGG16 ($sFuse$) or discriminator ($sFuseGAN$) respectively, uL1Loss ($uFuseGAN$) is the updated mean absolute error between generator output multiplied by the mask obtained by combining the masks of the input images and input images multiplied by corresponding masks.

\section{Results}
\label{sec:experiments}
\subsection{In-focus/out-of-focus images classification}
We created a dataset containing in-plain 45 specimens with all-in-focus images. We provide twenty levels of blur to every sample using z-stack frames from the video scans. For the evaluation, we select 100 focused and 100 unfocused crops, hand-labeled in a binary manner. To avoid random crops in the black area of the image, we create masks for each sample using binary threshold in OpenCV. We employ a binary classification task and accuracy metric (our test dataset is well-balanced for that). 

In our comparison, we consider the most significant focus measure operators used in microscopy~\cite{Redondo2011, Shah2017} (Laplacian, Tenengrad, Vollath), the original pre-trained model from the paper~\cite{Yang2019} and the proposed upgraded model. We also compare a popular pre-trained model for non-microscopy images from the work~\cite{Hsu2008}, based on the SVM model.





\begin{table}[h!]
\caption{ Accuracy comparison for focused classification task (CI = 95\%). The best results are highlighted in bold.}
\label{tab: in_out_focus}
\vskip 0.15in
\begin{center}
\begin{small}
\begin{sc}
\resizebox{\columnwidth}{!}{%
\begin{tabular}{|c|c|c|c|}

\hline

Focus measures    & SVM & Original model & Ours \\ \hline
\textless 55\% & 40\% &  $79\pm 2$\% & $\textbf{95,6} \pm \textbf{2,8\%}$ \\ \hline

\end{tabular}
}
\end{sc}
\end{small}
\end{center}
\vskip -0.1in
\end{table}

Results presented in Table~\ref{tab: in_out_focus} show that the proposed approach achieves better performance in terms of classification accuracy. Without fear of favor, the main reason of that is the fact that all considered methods are suited well for their specific tasks and perform poorly on other types of data. The main contribution of in-focus/out-of-focus classification task is the sustainability of the approach for false positives on the dusted images and artifacts.

\subsection{Fast scanning image deblurring}

\begin{figure}[htb]
\centering
\centerline{\includegraphics[width=\columnwidth]{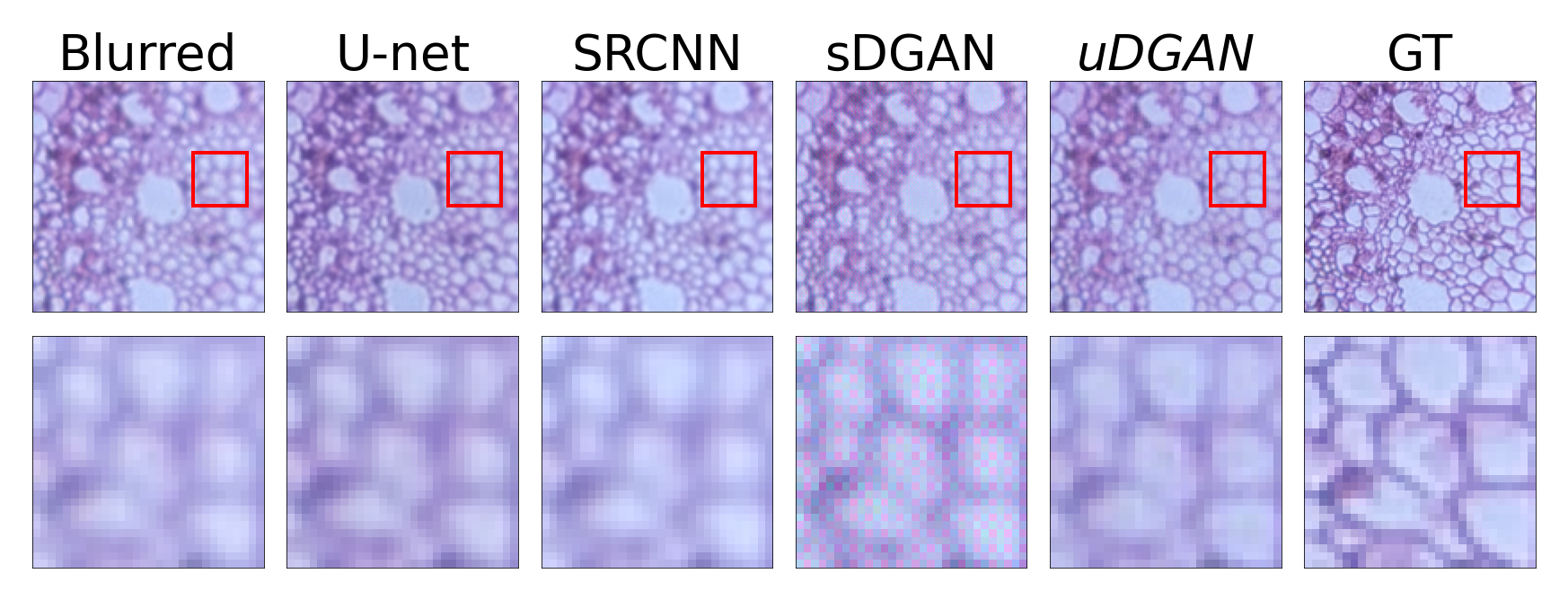}}
\caption{Comparison of deblurring models: U-net, CNN based and GAN based. Results of our modified model are highlighted in italic.}
\label{fig:dgan_res}
\end{figure}

To assess the quality of image restoration we employ structural similarity index (SSIM) and peak signal-to-noise ratio (PSNR) metrics. From the results presented in Table~\ref{tab: deblur_compar1} it can be seen that SRCNN provides better restoration quality of images in terms of PSNR and SSIM. In the meantime, Figure \ref{fig:dgan_res} shows that uDeblurGAN tends to provide sharper restoration of image details. Moreover, Figure~\ref{fig:dgan_res} shows that sDeblurGAN produces pictures with artifacts, which has the lowest PSNR and SSIM. We propose that these quality metrics could be used for rough estimation of performance~\cite{Ledig2017}, e.g. uDeblurGAN shows high PSNR and SSIM and generates images without artefacts because of the substitution of 2D convolution transpose layers from sDeblurGAN with resize-convolution layers.

\begin{table}[h!]
\caption{ Comparison of deblurring nets (mean over 8 test samples). Subscripts $s$ and $u$ denote standard and upgraded versions of DeblurGAN, respectively (CI = 95\%). The best results are highlighted in bold. Our modified model is highlighted in italic.}
\label{tab: deblur_compar1}
\vskip 0.15in
\begin{center}
\begin{small}
\begin{sc}
\begin{tabular}{|l|c|c|}
\hline
 & PSNR & SSIM, $10^{-2}$ \\ \hline
Blurred image      & $29.20\pm9.27$ & $86.88\pm 14.32$\\ \hline
U-net     & $26.11\pm4.19$ & $83.81\pm 10.21$\\ \hline
SRCNN     & $\textbf{28.80}\pm \textbf{7.83}$ & $\textbf{86.33}\pm \textbf{11.73}$\\ \hline
\emph{uDeblurGAN} & $28.33\pm8.09$ & $84.58\pm13.76$\\ \hline
sDeblurGAN & $24.49\pm 5.23$ & $62.54\pm18.26$\\ \hline
\end{tabular}
\end{sc}
\end{small}
\end{center}
\vskip -0.1in

\end{table}

\subsection{Focus-stacking}

\begin{figure}[htb]
\centering
\centerline{\includegraphics[width=\columnwidth]{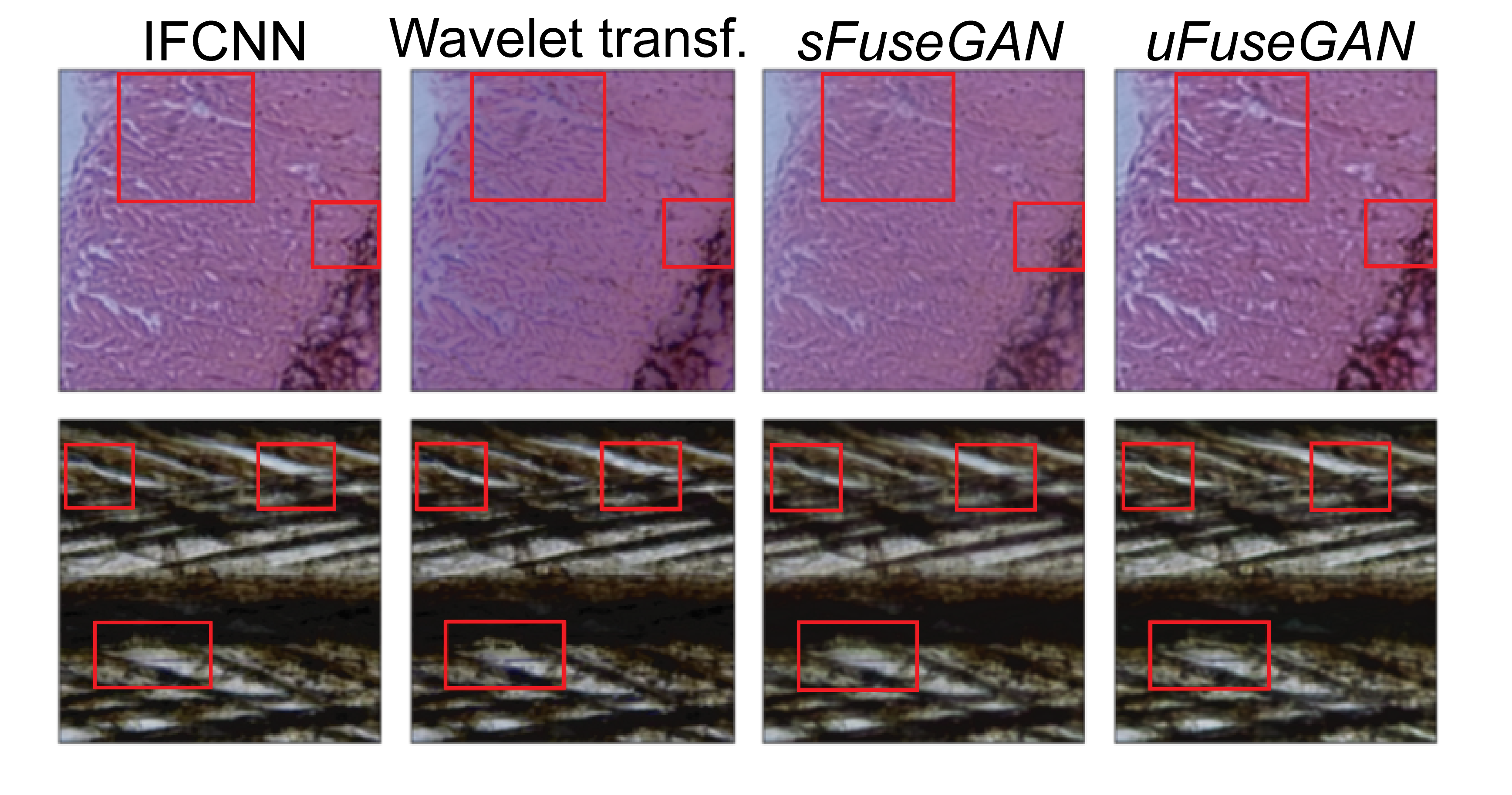}}
\caption{Comparison of several focus stacking approaches: CNN based, classical CV, and two GAN based. The red border highlights the main differences in quality.}
\label{fig:fstack_res}
\end{figure}

\begin{table}[t]
\caption{Performance of fusion models (mean over test samples). Subscripts $s$ and $u$ denote standard and upgraded versions of FuseGAN, respectively (CI = 95\%). The best results are highlighted in bold. Our modified models are highlighted in italic.}
\label{tab: quant comparison}
\vskip 0.15in
\begin{center}
\begin{small}
\begin{sc}
\begin{tabular}{|l|c|c|}
\hline
Model & BRISQUE & CNNIQA \\ \hline
$uFuseGAN$         & $37.21 \pm 3.48$ & $\textbf{25.71} \pm \textbf{4.36}$ \\ \hline
$sFuseGAN$         & $40.59 \pm 4.66$  & $32.44 \pm 5.44$ \\ \hline
Wavelet transform          & $\textbf{31.83} \pm \textbf{3.12}$   & $\textbf{23.47} \pm \textbf{3.01}$  \\ \hline
IFCNN  & $\textbf{36.19} \pm \textbf{3.23}$  & $25.97 \pm 4.48$  \\ \hline
\end{tabular}
\end{sc}
\end{small}
\end{center}
\vskip -0.1in
\end{table}

\noindent During the evaluation of the image fusion algorithms, we adopt both qualitative and quantitative methods to discriminate the performance of different image fusion algorithms. Firstly, qualitative evaluation is performed by judging the visual effects of their fusion images with respect to the source images. Since there is no ground-truth image available for that task, the pre-trained CNN (CNNIQA)~\cite{Kang2015} and the Blind/Referenceless Image Spatial QUality Evaluator (BRISQUE)~\cite{Mittal2012} are used for no-reference quantitative image quality assessment.

An anonymous survey, to determine the best algorithm for the obtained images, shows that qualitatively the best model is the upgraded FuseGAN model by dominant decision of 28 experts in bioimaging.
Experiments show that even though we are able to achieve qualitative improvements for the image fusion task, we are not able to improve the quantitative metrics that we have chosen to evaluate our models. Wavelets transform is a transform-based algorithm, i.e. they translate images to another domain, glue them together there and reverse the transformation. Therefore, the image turns out to be smooth with smooth transitions. Contrary, the proposed methods have pixel-based approach, so it can still get non-smooth transitions that are invisible to human eye and the metrics show that. Therefore, based on visual improvements, we can conclude that we successfully solve the image fusion task with proposed FuseGAN model.

\section{Conclusion}
\label{sec:conclusion}

In this work, we review existing deep learning based microscopy image processing methods and propose their modifications to solve relevant problems in mobile microscopy. The numerical comparison in the in-focus/out-of-focus classification task and qualitative comparison in the image deblurring and focus-stacking tasks with (both non-deep learning and deep learning based approaches) show that the proposed schemes achieve competitive results. We proposed a set of modules that allow to improve the acquired images and that could be stacked to a comprehensive pipeline, yielding a tuned solution for specific biomedical taks.
Realization of the end-to-end pipeline will be the subject of future work.


\section{Compliance with Ethical Standards}

This is a numerical simulation study on a custom dataset acquired specifically for this study. All regulatory ethical requirements have been met.
\section{Acknowledgments}
We would like to express our gratitude to Victor Lempitsky for helpful discussions and to the company MEL Science for providing the mobile microscopy dataset.
No funding was received for conducting this study. The authors disclose the absence of conflict of interests.

%


\bibliographystyle{IEEEbib}
\bibliography{refs}




\end{document}